\documentclass[a4paper,10pt]{article}
\usepackage{amsthm}
\usepackage{amsfonts}
\usepackage{amssymb}
\usepackage{mathrsfs}
\usepackage[english]{babel}
\usepackage{float}
\usepackage[toc,page]{appendix}
\usepackage{graphicx}
\usepackage{amsmath}
\usepackage{color}
\usepackage{setspace}

\usepackage{geometry}
\numberwithin{equation}{section}
\newtheorem{theorem}{Theorem}

\newtheorem{definition}[theorem]{Definition}

\DeclareGraphicsExtensions{.pdf,.png,.jpg}

\begin{document}

\title{Sparse Sampling in Helical Cone-Beam CT Perfect Reconstruction Algorithms}
\date{}
\author{Tamir Bendory and Arie Feuer}

\maketitle

\begin{abstract}
In the current paper we consider the Helical Cone Beam CT. This
scanning method exposes the patient to large quantities of radiation and results
in very large amounts of data being collected and stored. Both these facts are
prime motivators for the development of an efficient, reduced rate, sampling
pattern. 
We calculate bounds on the support in the frequency
domain of the collected data and use these to suggest an efficient sampling pattern. 
A reduction of up to a factor of {2} in sampling rate is suggested. Indeed, we
show that reconstruction quality is not affected by this reduction of sampling
rates. 
\end{abstract}

\section{Introduction}

In computerized tomography (CT) a patient is being scanned by a rotating
illuminating source (e.g. X-ray) and the projections of the rays
passing through the body are being recorded on an array of sensors. These
projections are then used to reconstruct either a cross section of the
scanned body (2D CT) or the whole scanned body (3D CT). Over the past decades
CT became very popular in medical imaging because it enables the physician
to visualize the internal structure of the patient, or at least some
properties of it (e.g. X-ray attenuation), without the need to actually
invade the patient interior.

In the current paper we consider the Helical Cone Beam (HCB) 3D CT. In this
system, while a cone beam X-ray source and a 2D array of sensors
are rotating around the scanned body, the bed with the body is moving along
the rotational axis. The result, from the scanned body point of view, is
equivalent to the X-ray source and sensor array moving along a helical path.
Clearly, this scanning method exposes the patient to large quantities of
radiation and results in very large amounts of data being collected and
stored. Both these facts are prime motivators for the development of an
efficient, reduced rate, sampling pattern which is the goal of the work
presented here.

It is known that theoretically, given the projections data, one can
reconstruct the scanned body in both the 2D and 3D cases. However, in practice, only
sampled versions of these projections are available. Clearly, if the
sampling is sufficiently dense, the body image can be reconstructed to any
desired quality. However, as sampling rate is closely associated with the
amount of radiation to which the scanned body is exposed, the challenge is
to sample as efficiently as possible while still maintaining a satisfactory
reconstructed image quality.

Dose reduction in CT has drawn a considerable attention recently. For instance, in \cite{fahimian2010low,Fahimian2,zhao2012high} it was shown that the equally sloped tomography algorithm can achieve good reconstruction quality with reduced x-ray dose. In the last decade, a sparsity prior has been used extensively to reduce the sampling rate in a variety of applications  (e.g. \cite{donoho2006compressed,Bar-IlanSub-Nyquis}) and to super-resolve signals \cite{candes2013towards,bendory2015super}.  
 Particularly, several attempts were made to reduce the sampling rate by compressed sensing methods \cite{6637820,hashemi2014efficient,shtok2013learned}.
In contrast, our work does not build upon a sparsity assumption and contributes to the ongoing effort by showing that the standard uniform sampling leads to significant redundancy. 

The common sampling used is uniform sampling. Rattey and Lindgran \cite{Sampling_2D_Radon} were the first to investigate the possibility of a more
efficient (i.e. reduced rate) sampling pattern for the parallel beam 2D CT.
They observed that the impulse response of the scanning system has a bow-tie
shape in the frequency domain and used the concept of essential bandwidth to
derive an interlaced sampling pattern of lower density, while maintaining the
quality of the reconstructed image. Natterer in \cite{Sampling_2D_fan} and Desbat et al. in \cite{Sampling_3D_fanbeam}
have generalized this result to the 2D fan beam case and 3D fan beam case,
respectively.

We generalize these works, calculate (essential) bounds on the support in the
frequency domain and use these to suggest an efficient sampling pattern for
the HCB CT. A reduction of up to a factor of {$2$} in sampling rate is
suggested. 
The sampling pattern involves a particular low-pass filtering  before sampling. In the presence of noise, this preliminary filtering will reduce the effects of measurement noise. This point will become clearer after the mathematical analysis of the following sections.

The outline of the paper is as follows. In Section 2 we introduce some
notation and the system geometry. In Section 3, which is the main part of this
paper, we analyze the frequency domain of the HCB scan and
calculate bounds on its essential support. In Section 4 we use this analysis
to suggest an efficient sampling pattern. In Section 5 we compare
reconstructions of a common phantom using our sampling scheme with the
standard sampling and finally, in Section 6, we summarize the results and
draw some conclusions.

\section{Notations, Geometry and Reconstruction}
Let us start with some notations. We use $\mathbf{x}=\left[ x,y,z\right]
^{T} $ for the Cartesian coordinate systems, '$\ast $' to denote linear
convolution and '$\left\lfloor \cdot \right\rfloor 
$' to denote the floor operation.

\begin{figure}[!]
  \centering
   \includegraphics[height=5in]{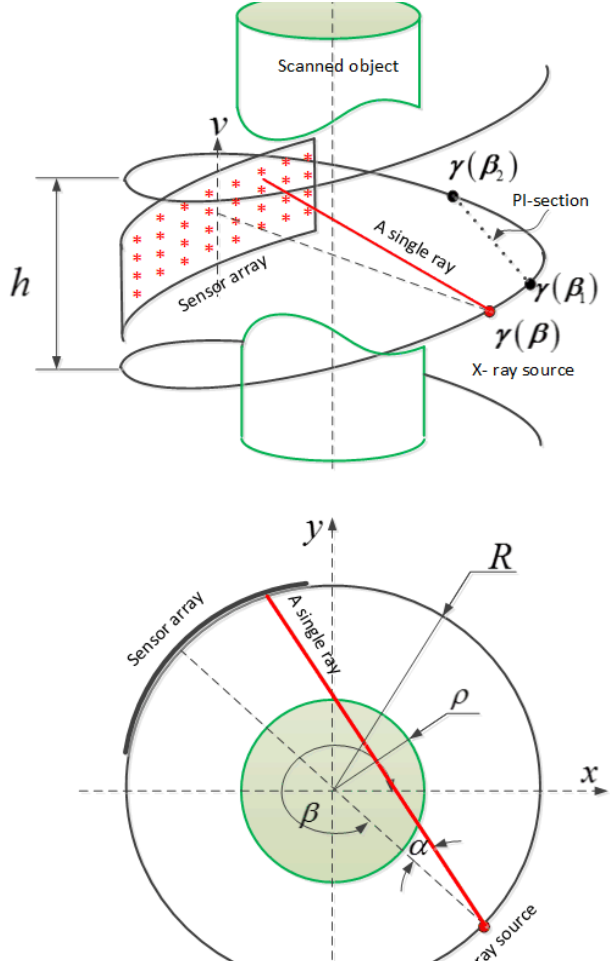}
  \caption{The helical scanning system configuration}
	\label{fig:helix}
\end{figure}

Let the function $f(\mathbf{x})$ be an unknown smooth attenuation function
which represents a patient to be scanned. We recall that the Fourier transform of the
function $f(\mathbf{x})$ is given by
\begin{equation}
\hat{f}(\mathbf{\omega })=\int_{R^{3}}f(\mathbf{x})e^{-j\mathbf{x}^{T}%
\mathbf{\omega }}d\mathbf{x}.  \label{1}
\end{equation}%
The function $f(\mathbf{x})$ is confined to a cylinder of radius $\rho $ and
is scanned by a scanner which rotates at radius $R>\rho >0$. The scanned
patient bed is moving along the $z$-axis at a constant rate. The result is
equivalent to having the scanned body stationary and the scanning system
moving on a helical trajectory defined by the set 
\begin{equation}
\Gamma (\beta )=\left\{ \,\mathbf{\gamma }(\beta )=\left[ R\cos {\beta }%
,R\sin {\beta },{\frac{h}{2\pi }}\beta \right] ^{T}|\,\beta \in \lbrack
-B,B]\right\},  \label{2}
\end{equation}
where we assume that the scanned body is of finite length, $z\in[-Z,Z]$, resulting in a finite range for $\beta\in[-B,B]$,
and $h$ denotes a constant helix pitch \footnote{
Pitch is the distance along the helix's axis that is covered by one complete
rotation of the helix (see Figure \ref{fig:helix}).}. A schematic layout of
the HCB CT is presented in Figure \ref{fig:helix}. 

\begin{definition}
\label{definition1} \bigskip Let $\mathbf{\gamma }\left( \beta _{1}\right) $, 
$\mathbf{\gamma }\left( \beta _{2}\right) \in \Gamma (\beta )$ be such that $
\left\vert \beta _{1}-\beta _{2}\right\vert <2\pi $. The line 
\begin{equation}
\mathbf{x}\left( \beta _{1},\beta _{2}\right) =\left( 1-t\right) \mathbf{
\gamma }\left( \beta _{1}\right) +t\mathbf{\gamma }\left( \beta _{2}\right), 
\quad t\in \mathbb{R},  \label{3}
\end{equation}
is commonly referred to as a \emph{PI-line}. When one restricts $0\leq t\leq
1$ one gets a PI-section (see Figure \ref{fig:helix}).
\end{definition}

With Definition \ref{definition1} we can quote an interesting geometrical property \cite{Solution_to_long_object}:

\begin{theorem}
\label{theorem1}(Helix Property) Consider the helix trajectory as defined in
(\ref{2}). Then every point $\mathbf{x}$ confined within the helix has a unique PI-line passing
through it.
\end{theorem}

An immediate consequence of Theorem \ref{theorem1} is the following
coordinate transform:
\begin{align*}
x& =R[(1-t)\cos {\beta_{1}}+t\cos {\beta_{2}}] ,
\label{eq:coordinate transformation} \\ \nonumber
y& =R[(1-t)\sin {\beta_{1}}+t\sin {\beta_{2}}]  ,  \\ \nonumber 
z& =\frac{h}{2\pi }[(1-t)\beta_{1}+t\beta_{2}] , \nonumber 
\end{align*}
where $t\in \lbrack 0,1]$, which is a one-to-one transform provided $
\left\vert \beta _{1}-\beta _{2}\right\vert <2\pi $.

The data collected is the set of all projections of rays going through the
scanned body. Specifically, a projection is the integral along a ray
emanating from the source at position $\mathbf{\gamma }(\beta )$, passing
through the scanned object $f(\mathbf{x})$ and hitting the sensors array at
the coordinates $\left[ \alpha ,v\right] ^{T}$. The
aim of a reconstruction algorithm is to reconstruct $f(\mathbf{x})$ from its
projections, namely, the measured data. The (continuous) projection data is
given by%
\begin{equation}
\mathcal{D}f\left( \alpha ,\beta ,v\right) =\int_{\mathbf{x}\left( \alpha
,\beta ,v\right) }f\left( \mathbf{x}\right) d\mathbf{x},  \label{5}
\end{equation}
where by $\mathbf{x}\left( \alpha ,\beta ,v\right) $ we denote all the
points on the line (ray) connecting $\mathbf{\gamma }(\beta )$ to $\left[
\alpha ,v\right] ^{T}$ on the sensor array.

Before we conclude this section, we wish to point out that there are a number
of perfect reconstruction algorithms using the data $\mathcal{D}f\left(
\alpha ,\beta ,v\right) $. That is to say that one can reconstruct the exact
image, $f(\mathbf{x})$, from its helical cone beam projections. Good
examples are found in \cite{Katsevich,Zou_Pan} which are of the filtered
back-projection type and have computational advantages. These algorithms use
the Helix property to reconstruct each PI-section independently using data
corresponding to $\beta \in \lbrack \beta _{1},\beta _{2}]$.

\section{Essential Support Analysis}

While perfect reconstruction of $f\left( \mathbf{x}\right) $ from $\mathcal{D%
}f\left( \alpha ,\beta ,v\right) $ is theoretically possible, in practice,
one can only collect a sampled version of $\mathcal{D}f\left( \alpha ,\beta
,v\right) $. The commonly used sampling is uniform in $\left( \alpha ,\beta
,v\right) $ and to avoid aliasing it needs to be sufficiently dense (in the sense of Nyquist
rate). This, however, as pointed out earlier, results in patient exposure\
to high amounts of radiation and in very large amounts of data which needs
to be stored and processed. To alleviate some of these problems, we will
investigate the frequency content of $\mathcal{D}f\left( \alpha ,\beta
,v\right)$, and adapt the sampling pattern to its support. Strictly
speaking, there are no band-limited functions of finite spatial support.
Thus, we use the term essential support and work with functions that have a
finite essential support in the frequency domain.

\begin{definition}
\label{definition2}Given the function $g(\mathbf{\omega }),$ $\mathbf{\omega
\in }\mathbb{R}^{3}$, its essential support is given by a set $\mathcal{L}%
\left\{ g(\mathbf{\omega })\right\} \subset \mathbb{R}^{3}$ such that 
\begin{equation}
\int_{\mathcal{L}\left\{ g(\mathbf{\omega })\right\} }\left\vert g(\mathbf{%
\omega })\right\vert ^{2}d\mathbf{\omega }\geq 0.98\int_{\mathbb{R}%
^{3}}\left\vert g(\mathbf{\omega })\right\vert ^{2}d\mathbf{\omega }.
\label{6}
\end{equation}
\end{definition}

Since the function we consider is $\mathcal{D}f\left( \alpha ,\beta
,v\right) $ and it is periodic in $\alpha $, its Fourier transform will have
the form%
\begin{equation}
\widehat{\mathcal{D}f}\left( \omega _{\alpha },\omega _{\beta },\omega
_{v}\right) =\sum_{m\in \mathbb{Z}}\widehat{\mathcal{D}f}\left( m,\omega
_{\beta },\omega _{v}\right) \delta \left( \omega _{\alpha }-m\right),
\label{6a}
\end{equation}%
where%
\begin{equation*}
\widehat{\mathcal{D}f}\left( m,\omega _{\beta },\omega _{v}\right) =\frac{1}{%
2\pi }\int_{-\pi }^{\pi }d\alpha \int_{-\infty }^{\infty
}dv\int_{-B}^{B}d\beta \mathcal{D}f\left( \alpha ,\beta ,v\right)
e^{-j\left( \alpha m+v\omega _{v}+\beta \omega _{\beta }\right) }.  \label{7}
\end{equation*}%
Substituting (\ref{5}), we get%
\begin{equation}
\widehat{\mathcal{D}f}\left( m,\omega _{\beta },\omega _{v}\right) =\frac{1%
}{2\pi }\int_{-\pi }^{\pi }d\alpha \int_{-\infty }^{\infty
}dv\int_{-B}^{B}d\beta \int_{\mathbf{x}\left( \alpha ,\beta ,v\right)
}f\left( \mathbf{x}\right) d\mathbf{x}e^{-j\left( \alpha m+v\omega
_{v}+\beta \omega _{\beta }\right) }. 
 \label{8}
\end{equation}

We start by observing from Figure \ref{fig:helix} that 
\begin{equation*}
\begin{split}
\mathbf{x}\left( \alpha ,\beta ,v\right)&=\left\{ \mathbf{x\in }\mathbb{R}
^{3}:\left( R-x\cos \beta -y\sin \beta \right) \sin \alpha =\left( x\sin
\beta -y\cos \beta \right) \cos \alpha \right. ,\\
&\left. v\left( \left( R-x\cos \beta -y\sin \beta \right) \cos a+\sin
\alpha \left( x\sin \beta -y\cos \beta \right) \right) =2R\cos \alpha \left(
z-\frac{h}{2\pi }\beta \right) \right\}.
\end{split}
\end{equation*}
So, we can rewrite (\ref{8}) as%
\begin{equation}
\begin{split}
&\widehat{\mathcal{D}f}\left( m,\omega _{\beta },\omega _{v}\right)=\frac{1
}{2\pi }\int_{\mathbb{R}^{3}}d\mathbf{x}f\left( \mathbf{x}\right) \int_{-\pi
}^{\pi }d\alpha \int_{-\infty }^{\infty }dv\int_{-B}^{B}d\beta e^{-j\left(
\alpha m+v\omega _{v}+\beta \omega _{\beta }\right) }  \\
&\delta \left( \left( R-x\cos \beta -y\sin \beta \right) \sin \alpha
-\left( x\sin \beta -y\cos \beta \right) \cos \alpha \right)  \\
&\delta \left( v\left( \left( R-x\cos \beta -y\sin \beta \right) \cos
\alpha +\sin \alpha \left( x\sin \beta -y\cos \beta \right) \right) -2R\cos
\alpha \left( z-\frac{h}{2\pi }\beta \right) \right)  \\
&=\int_{\mathbb{R}^{3}}d\mathbf{x}f\left( \mathbf{x}\right) E_{\mathbf{x}%
}\left( m,\omega _{\beta },\omega _{v}\right). 
\end{split} \label{9}
\end{equation}

Our goal will now be to find $\mathcal{E}_{\mathbf{x}}=\mathcal{L}\left\{ E_{%
\mathbf{x}}\left( m,\omega _{\beta },\omega _{v}\right) \right\} $ and then
find the minimal set $\mathcal{E}$ such that 
\begin{equation}
\mathcal{E}_{\mathbf{x}} \subseteq \mathcal{E}, \quad \left\{\mathbf{x}=\left[ x,y,z\right]\thinspace : \thinspace x^{2}+y^{2}\leq \rho\thinspace ,\thinspace \vert z\vert \leq Z\right\}. \label{9a}
\end{equation}
Clearly, by (\ref{6a}) and (\ref{9}), $\mathcal{L}\left\{\widehat{\mathcal{D}f}\left( \omega_\alpha,\omega _{\beta
},\omega _{v}\right)\right\}\subset \mathcal{E}$.
Using properties of Delta functions and (\ref{9}), we can write 
\begin{equation}
\begin{split}
E_{\mathbf{x}}\left( m,\omega _{\beta },\omega _{v}\right)&=\frac{1}{2\pi }
\int_{-\pi }^{\pi }d\alpha \int_{-\infty }^{\infty }dv\int_{-\infty
}^{\infty }d\beta e^{-j\left( \alpha m+v\omega _{v}+\beta \omega _{\beta
}\right) }   \\
&=g_{1}\left( \alpha ,\beta \right) g_{2}\left( \beta ,v\right),
\end{split}\label{10}
\end{equation}
where
\begin{align*}
g_{1}\left( \alpha ,\beta \right)&:=\delta \left( \left( R-x\cos \beta
-y\sin \beta \right) \sin \alpha -\left( x\sin \beta -y\cos \beta \right)
\cos \alpha \right),  \notag \\
g_{2}\left( \beta ,v\right) &:=\frac{I_{B}\left( \beta \right) }{{%
\left( R-x\cos \beta -y\sin \beta \right) ^{2}+\left( x\sin \beta -y\cos
\beta \right) ^{2}}}  \notag \\
&\delta \left( v  -2R\left( z-\frac{h}{2\pi }\beta \right) \frac{R-x\cos \beta -y\sin
\beta }{{\left( R-x\cos \beta -y\sin \beta \right)^{2}+\left( x\sin
\beta -y\cos \beta \right) ^{2}}}\right),  \label{100}
\end{align*}
where $I_{B}\left( \beta \right) $ is the indicator function for the
interval $\left[ -B,B\right] $. Rewriting (\ref{10}) we then have
\begin{equation}
\begin{split}
E_{\mathbf{x}}\left( m,\omega _{\beta },\omega _{v}\right)&= 
\frac{1}{2\pi }\int_{-\infty }^{\infty }d\beta e^{-j\beta \omega _{\beta }}\int_{-\pi
}^{\pi }d\alpha e^{-j\alpha m}g_{1}\left( \alpha ,\beta \right) 
\int_{-\infty }^{\infty }dve^{-jv\omega _{v}}g_{2}\left( \beta ,v\right) 
 \\
&=\widehat{g}_{1}\left( m,\omega _{\beta }\right) \ast \widehat{g}
_{2}\left( \omega _{\beta },\omega _{v}\right),  \label{101}
\end{split}
\end{equation}
where the mono-dimensional convolution is with respect to the variable $\omega _{\beta }$ and 
\begin{equation*}
\widehat{g}_{1}\left( m,\omega _{\beta }\right) =\frac{1}{2\pi }%
\int_{-\infty }^{\infty }\int_{-\pi }^{\pi }d\alpha d\beta e^{-j\beta \omega
_{\beta }}e^{-j\alpha m}g_{1}\left( \alpha ,\beta \right),  \label{103}
\end{equation*}%
\begin{equation*}
\widehat{g}_{2}\left( \omega _{\beta },\omega _{v}\right) =\int_{-\infty
}^{\infty }\int_{-\infty }^{\infty }dvd\beta e^{-j\beta \omega _{\beta
}}e^{-jv\omega _{v}}g_{2}\left( \beta ,v\right) . \label{104}
\end{equation*}%
Since $g_{1}\left( \alpha ,\beta \right) $ is periodic in $\beta $, we have%
\begin{equation*}
\widehat{g}_{1}\left( m,\omega _{\beta }\right) =\sum_{k\in \mathbb{Z}}%
\widehat{g}_{1}\left( m,k\right) \delta \left( \omega _{\beta }-k\right),
\end{equation*}%
where%
\begin{equation}
\widehat{g}_{1}\left( m,k\right) =\frac{1}{\left( 2\pi \right) ^{2}}%
\int_{-\pi }^{\pi }\int_{-\pi }^{\pi }d\alpha d\beta e^{-j\left( \alpha
m+\beta k\right) }g_{1}\left( \alpha ,\beta \right) . \label{105}
\end{equation}%
The integral in (\ref{105}) is the one investigated by Natterer \cite{Sampling_2D_fan}, where
the essential support, under the assumption that $f\left( \mathbf{x}\right) $%
\ is essentially band limited to a ball of radius $\Omega $, was found to be%
\begin{equation}
\mathcal{L}\left\{ \widehat{g}_{1}\left( m,k\right) \right\} =\left\{ \left(
m,k\right) :\left\vert k-m\right\vert <\Omega R,\left\vert k\right\vert
<\left\vert k-m\right\vert \rho \right\} . \label{106}
\end{equation}
\bigskip Let us now concentrate on calculating the essential support of $%
\widehat{g}_{2}\left( \omega _{\beta },\omega _{v}\right) $. We first
introduce the notations
\begin{align*}
\overline{r} &=\frac{\sqrt{x^{2}+y^{2}}}{R} , \notag \\
\phi &=\arctan \left( \frac{y}{x}\right) . \label{11}
\end{align*}
Then, $g_{2}\left( \beta ,v\right) $ becomes
\begin{equation}
\begin{split}
g_{2}\left( \beta ,v\right) &=\frac{I_{B}\left( \beta \right) }{R^2\left(1+
\overline{r}^{2}-2\overline{r}\cos \left( \beta -\phi \right)\right)} \\
&\delta \left( v-2\left( z-\frac{h}{2\pi }\beta \right) \frac{1-\overline{r}\cos
\left( \beta -\phi \right) }{{1+\overline{r}^{2}-2\overline{r}\cos
\left( \beta -\phi \right) }}\right),  \label{12}
\end{split}
\end{equation}
and, substituting (\ref{12}) we get%
\begin{equation*}
\begin{split}
\widehat{\tilde{g}}_{2}\left( \beta ,\omega _{v}\right) &=\int_{-\infty }^{\infty
}dve^{-jv\omega _{v}}g_{2}\left( \beta ,v\right) \\
&=\frac{I_{B}\left( \beta \right) }{R^{2}\left( 1+\overline{r}^{2}-2
\overline{r}\cos \left( \beta -\phi \right) \right) }e^{-j\frac{2\omega
_{v}\left( z-\frac{h}{2\pi }\beta \right) \left( 1-\overline{r}\cos \left(
\beta -\phi \right) \right) }{1+\overline{r}^{2}-2\overline{r}\cos \left(
\beta -\phi \right) }} , \label{13}
\end{split}
\end{equation*}
and%
\begin{equation*}
\begin{split}
\widehat{g}_{2}\left( \omega _{\beta },\omega _{v}\right) &=\int_{-\infty
}^{\infty }d\beta e^{-j\beta \omega _{\beta }}\widehat{\tilde{g}}_{2}\left( \beta
,\omega _{v}\right)   \\
&=\frac{1}{R^{2}}\int_{-\infty }^{\infty }d\beta e^{-j\beta \omega _{\beta
}}g_{3}\left( \beta -\phi \right) I_{B}\left( \beta \right) g_{4}\left(
\beta ,\omega _{v}\right),  \label{14}
\end{split}
\end{equation*}
where%
\begin{align}
g_{3}\left( \beta \right) &:=\frac{1}{1+\overline{r}^{2}-2\overline{r}\cos
\beta } , \notag \\
g_{4}\left( \beta ,\omega _{v}\right) &=:e^{-j2\omega _{v}\left( z-\frac{h}{%
2\pi }\beta \right) I_{B}\left( \beta \right) g_{5}\left( \beta -\phi
\right) },  \notag \\
g_{5}\left( \beta \right) &=:\frac{1-\overline{r}\cos \beta }{1+\overline{r}%
^{2}-2\overline{r}\cos \beta }.  \label{15}
\end{align}
In Appendix A, we calculate the constraints for the essential supports of $%
\widehat{g}_{3}\left( \omega _{\beta }\right) $ and $\widehat{g}_{4}\left(
\omega _{\beta },\omega _{v}\right) $ to be
\begin{align*}
\mathcal{L}\left\{ \widehat{g}_{3}\left( \omega _{\beta }\right) \right\}
&=\left\{ \omega _{\beta }:\left\vert \omega _{\beta }\right\vert \leq
K\right\},  \notag \\
\mathcal{L}\left\{ \widehat{g}_{4}\left( \omega _{\beta },\omega _{v}\right)
\right\} &\subset\left\{ \omega _{\beta }:\left\vert \omega _{\beta
}\right\vert \leq \left( K+\frac{48}{B}\right) \left( 1+C\left( z,\omega
_{v}\right) \right) \right\},  
\end{align*}
where%
\begin{align*}
K &=\left\lfloor \frac{\log \left( 0.01\left( 1+\overline{r}^{2}\right)
\right) }{2\log \overline{r}}\right\rfloor , \notag \\
C\left( z,\omega _{v}\right) &=\frac{2\left\vert \omega _{v}\right\vert }{1-%
\overline{r}}\left( \left\vert z\right\vert +\frac{h}{2\pi }B\right).
\label{52}
\end{align*}
Since the essential support of $\mathcal{L}\left\{ \widehat{I}_{B}\left(
\omega _{\beta }\right) \right\} =\left\{ \omega _{\beta }:\left\vert \omega
_{\beta }\right\vert \leq \frac{16}{B}\right\} $, we get%
\begin{equation*}
\begin{split}
\mathcal{L}\left\{ \widehat{g}_{2}\left( \omega _{\beta },\omega _{v}\right)
\right\} &\subset \mathcal{L}\left\{ \widehat{g}_{3}\left( \omega _{\beta
}\right) \right\} +\mathcal{L}\left\{ \widehat{I}_{B}\left( \omega _{\beta
}\right) \right\} +\mathcal{L}\left\{ \widehat{g}_{4}\left( \omega _{\beta
}\right) \right\}  \\
&=\left\{ \omega _{\beta }:\left\vert \omega _{\beta }\right\vert \leq K+%
\frac{16}{B}+\left( K+\frac{48}{B}\right) \left( 1+C\left( z,\omega
_{v}\right) \right) \right\} . \label{53}
\end{split}
\end{equation*}

So far the assumption that $f\left( \mathbf{x}\right) $ is essentially
band limited to the ball of radius $\Omega $ has been only used to generate (%
\ref{106}). We will now use it to derive an additional constraint. We start by an alternative description of the line $\mathbf{x}\left( \alpha
,\beta ,v\right) $%
\begin{equation*}
\begin{split}
&\mathbf{x}\left( \alpha ,\beta ,v\right) =\left\{ \mathbf{x\in }\mathbb{R} ^{3}: x=R\cos \beta -\widetilde{r}\cos (\alpha +\beta ) ,\right. \\
&\left.  y=R\sin \beta -\widetilde{r}\sin \left( \alpha +\beta \right)
, z=\frac{\widetilde{r}}{2R\cos \alpha }v+\frac{h}{2\pi }\beta ,%
 \widetilde{r}\mathbf{\in }\mathbb{R}\right\} . \label{54}
\end{split}
\end{equation*}
Then, from (\ref{1}) and (\ref{5}) we get%
\begin{equation*}
\begin{split}
\widehat{\mathcal{D}f}\left( \omega _{\alpha },\omega _{\beta },\omega
_{v}\right) &=\frac{1}{2\pi }\int_{-\pi }^{\pi }d\alpha e^{-j\omega_\alpha\alpha
}\int_{-B}^{B}d\beta e^{-j\beta \omega _{\beta }}\int_{\mathbb{R}
}dve^{-jv\omega _{v}} \\
&\int_{\mathbb{R}}f\left( R\cos \beta -\widetilde{r}\cos (\alpha +\beta
), R\sin \beta -\widetilde{r}\sin \left( \alpha +\beta \right) ,
\frac{\widetilde{r}}{2R\cos \alpha }v+\frac{h}{2\pi }\beta \right) d\widetilde{r},
\end{split}
\end{equation*}
and%
\begin{equation*}
\begin{split}
\widehat{\mathcal{D}f}\left( \omega _{\alpha },\omega _{\beta },\omega
_{v}\right) &=\frac{1}{\left( 2\pi \right) ^{4}}\int_{\mathbb{R}%
^{3}}d\omega _{x}d\omega _{y}d\omega _{z}\widehat{f}\left( \omega
_{x},\omega _{y},\omega _{z}\right)\int_{\mathbb{R}}d\widetilde{r} \int_{-\pi }^{\pi }d\alpha e^{-j\omega_\alpha\alpha }  \\
&\int_{-B}^{B}d\beta e^{-j\beta
\omega _{\beta }}e^{j\omega _{z}\frac{h}{2\pi }\beta }  
e^{j\omega _{x}\left( R\cos \beta -\widetilde{r}\cos (\alpha +\beta
\right) }e^{j\omega _{y}\left( R\sin \beta -\widetilde{r}\sin \left( \alpha
+\beta \right) \right) }  \\
&\int_{\mathbb{R}}dve^{-jv\left( \omega _{v}-\omega _{z}\frac{\widetilde{r}%
}{2R\cos \alpha }\right) }  .\label{55}
\end{split}
\end{equation*}
As $\int_{\mathbb{R}}dve^{-jv\left( \omega _{v}-\omega _{z}\frac{\widetilde{r%
}}{2R\cos \alpha }\right) }=2\pi \delta \left( \omega _{v}-\omega _{z}\frac{%
\widetilde{r}}{2R\cos \alpha }\right) $ and as we assume that $\widehat{f}%
\left( \omega _{x},\omega _{y},\omega _{z}\right) $ is (essentially) zero
for $\omega _{x}^{2}+\omega _{y}^{2}+\omega _{z}^{2}>\Omega ^{2}$ we can
readily conclude that $\widehat{\mathcal{D}f}\left( \omega _{\alpha },\omega
_{\beta },\omega _{v}\right) $ is zero for $\left\vert \omega
_{v}\right\vert >\underset{\widetilde{r},\alpha }{\max }\left\vert \frac{%
\widetilde{r}}{2R\cos \alpha }\right\vert \Omega $. Observing from Figure %
\ref{fig:helix} that $R-\rho \leq \widetilde{r}\leq R+\rho $ and $\left\vert
\alpha \right\vert \leq \arcsin \left( \frac{\rho }{R}\right) $ we conclude
that $\underset{\widetilde{r},\alpha }{\max }\left\vert \frac{\widetilde{r}}{%
2R\cos \alpha }\right\vert =\frac{R+\rho }{2\sqrt{R^{2}-\rho ^{2}}}$ and the
support of $\widehat{\mathcal{D}f}\left( \omega _{\alpha },\omega _{\beta
},\omega _{v}\right) $ is zero for%
\begin{equation}
\left\vert \omega _{v}\right\vert >\frac{R+\rho }{2\sqrt{R^{2}-\rho ^{2}}}\Omega =W_{v} 
 \label{56}.
\end{equation}

\begin{figure}[!]
  \centering
    \includegraphics[width=0.5\textwidth]{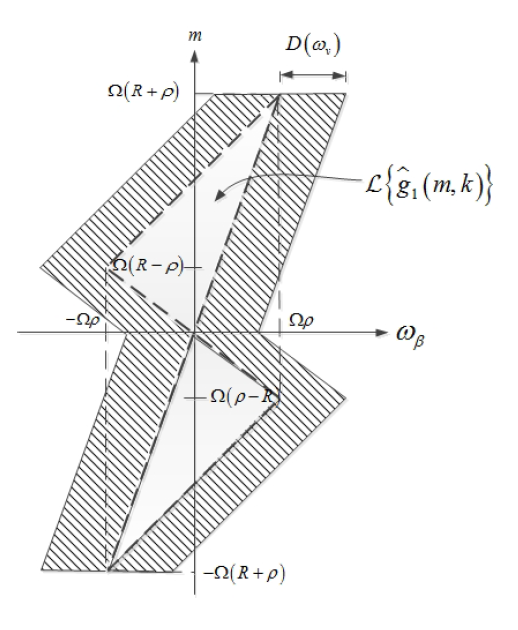}
  \caption{Cross section of the constrains set at $\omega_v$}
	\label{fig:2D_slice}
\end{figure}

We are ready now to combine the constraints we derived to get $\mathcal{E}$ as
defined in (\ref{9a}). We first rewrite more explicitly the set defined by (%
\ref{106}) (see Figure \ref{fig:2D_slice}),%
\begin{equation}
\begin{split}
&\mathcal{L}\left\{ \widehat{g}_{1}\left( m,k\right) \right\}=\left\{
\left( m,k\right) :\text{ s.t.}\right.  \\
&\left. 
\begin{array}[4]{llll}
m-\Omega R\leq k\leq \frac{\rho }{\rho +R}m\text{ \ for }\left( R-\rho
\right) \Omega \leq m\leq \left( \rho +R\right) \Omega \\ 
\frac{\rho }{\rho -R}m\leq k\leq \frac{\rho }{\rho +R}m\text{ \ for }0\leq
m\leq \left( R-\rho \right) \Omega \\ 
\frac{\rho }{\rho +R}m\leq k\leq \frac{\rho }{\rho -R}m\text{ \ for }\left(
\rho -R\right) \Omega \leq m\leq 0 \\ 
\frac{\rho }{\rho +R}m\leq k\leq m+\Omega R\text{ \ for }-\left( R+\rho
\right) \Omega \leq m\leq \left( \rho -R\right) \Omega%
\end{array}%
\right\}  \label{57}
\end{split}
\end{equation}

Observing now (\ref{101}) we conclude that the resulting support
of $E_{\mathbf{x}}\left( m,\omega _{\beta },\omega _{v}\right) $, will, for
each value of $\omega _{v}$ and $z$, be contained in the set in (\ref{57})
extended along the $\omega _{\beta }$ axis by the ($\omega _{v },z$)
dependent bound defined in (\ref{53}). So, if we denote 
\begin{equation}
D\left( \omega _{v}\right) =K+\frac{16}{B}+\left( K+\frac{48}{B}\right)
\left( 1+C\left( Z,\omega _{v}\right) \right) , \label{58}
\end{equation}%
we conclude by its definition that%
\begin{equation}
\begin{split}
&\mathcal{E} \mathcal{\subset }\left\{ \left( m,\omega _{\beta },\omega
_{v}\right) :\text{ s.t.}\right.  \\
&\left. 
\begin{array}{lllll}
m-\Omega R-D\left( \omega _{v}\right) \leq k\leq \frac{\rho }{\rho +R}%
m+D\left( \omega _{v}\right) \text{ \ for }\left( R-\rho \right) \Omega \leq
m\leq \left( \rho +R\right) \Omega \\ 
\frac{\rho }{\rho -R}m-D\left( \omega _{v}\right) \leq k\leq \frac{\rho }{%
\rho +R}m+D\left( \omega _{v}\right) \text{ \ for }0\leq m\leq \left( R-\rho
\right) \Omega \\ 
\frac{\rho }{\rho +R}m-D\left( \omega _{v}\right) \leq k\leq \frac{\rho }{%
\rho -R}m+D\left( \omega _{v}\right) \text{ \ for }\left( \rho -R\right)
\Omega \leq m\leq 0 \\ 
\frac{\rho }{\rho +R}m-D\left( \omega _{v}\right) \leq k\leq m+\Omega
R+D\left( \omega _{v}\right) \text{ \ for }-\left( R+\rho \right) \Omega
\leq m\leq \left( \rho -R\right) \Omega \\ 
\left\vert \omega _{v}\right\vert \leq \frac{R+\rho }{2\sqrt{R^{2}-\rho ^{2}}%
}\Omega%
\end{array}%
\right\}  \label{59}
\end{split}
\end{equation}
In Figure \ref{fig:2D_slice} we present a cross section of this set for a general
value of $\omega _{v}$.

\section{Efficient Sampling}

\begin{figure}[!]
  \centering
    \includegraphics[width=0.5\textwidth]{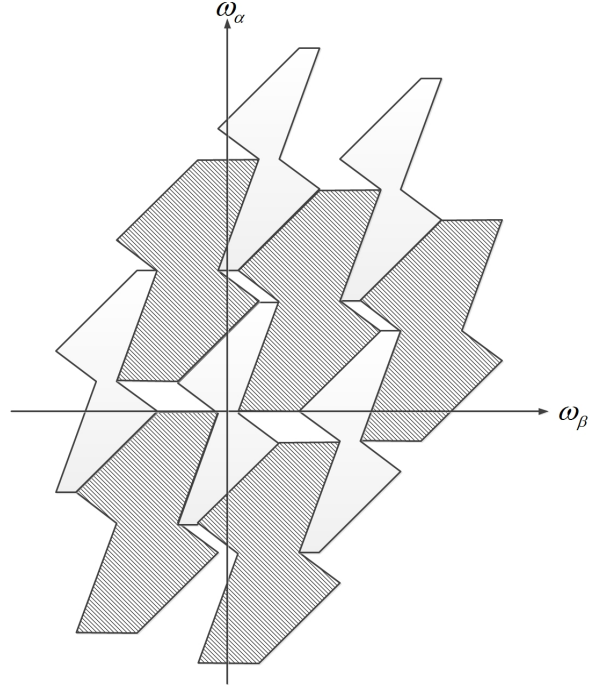}
  \caption{Tiling in the ($\omega_\beta$,$\omega_\alpha$) plane.}
	\label{fig:2D_tiling}
\end{figure}

We now use the set in (\ref{59}) to tile the 3D frequency space. To do that we consider a lattice in this 3D space. Then, around each lattice point we put a replica of the set (\ref{59}) and attempt to cover the 3D space with these replicas without them overlapping each other.   
The tightest tilling (by this we mean the least space uncovered by this tiling) was achieved by using the lattice defined by
\begin{equation}
\mathcal{L}at\left( 2\pi T^{-T}\right) =\left\{ \mathbf{\omega }\in \mathbb{R%
}^{3}:\mathbf{\omega =}2\pi T^{-T}\mathbf{n,n}\in \mathbb{Z}^{3}\right\},
\label{60}
\end{equation}%
where $T$ is the sampling matrix given by (see Figures \ref{fig:2D_tiling} and \ref%
{fig:3D_tiling})

\begin{figure}[!]
  \centering
    \includegraphics[width=0.5\textwidth]{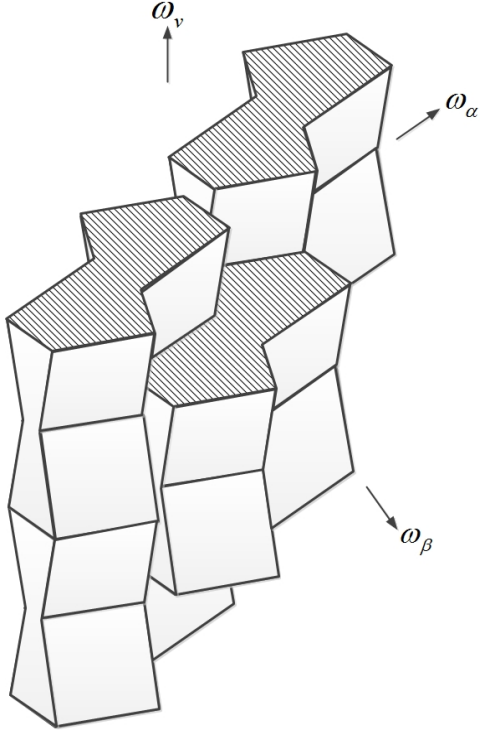}
  \caption{The tiling in the 3D frequency space.}
	\label{fig:3D_tiling}
\end{figure}
 
\begin{eqnarray}
&&2\pi T^{-T} =\left[ 
\begin{array}{ccc}
\Omega \left( R+\rho \right) & 2\Omega R & -2\Omega R \\ 
\Omega \rho +D & -D & D \\ 
W_{v} & W_{v} & W_{v}%
\end{array}%
\right]  \notag \\
&\Rightarrow &T=\frac{\pi}{W_v\Omega \left( 2R\Omega \rho +3RD+\rho
D\right) }  \notag \\
&&\left[ 
\begin{array}{ccc}
2DW_{v} & \Omega \rho W_{v} & -W_{v}\left( 2D+\Omega \rho \right) \\ 
4R\Omega W_{v} & -\Omega W_{v}\left( 3R+\rho \right) & \Omega W_{v}\left(
\rho -R\right) \\ 
0 & \Omega \left( 3RD+\rho D+2R\Omega \rho \right) & \Omega \left( 3RD+\rho
D+2R\Omega \rho \right)%
\end{array}%
\right]  \label{61}
\end{eqnarray}%
where 
\begin{equation*}
D=D\left( 0\right) +D\left( W_{v}\right),
\end{equation*}%
with $W_{v}$ and $D\left( \omega _{v}\right) $ defined in (\ref{56}) and (%
\ref{58}), respectively. So, the suggested sampling lattice is given by%
\begin{equation}
\mathcal{L}at\left( T\right) =\left\{ \left( \alpha ,\beta ,v\right) \in 
\mathbb{R}^{3}:\left( \alpha ,\beta ,v\right) ^{T}\mathbf{=}T\mathbf{k,k}\in 
\mathbb{Z}^{3}\right\} . \label{63}
\end{equation}

The common alternative to the suggested sampling is the uniform sampling
corresponding to the set in (\ref{58}). The bounding box in the frequency
domain is given by (see Figure \ref{fig:2D_slice}) $\left\vert \omega
_{\alpha}\right\vert \leq \Omega \left( R+\rho \right) $, $\left\vert \omega
_{\beta }\right\vert \leq \Omega\rho  +D\left(
W_{v}\right) ,\left\vert \omega _{v}\right\vert \leq W_{v}$. Namely,
the sampling matrix is given by%
\begin{equation*}
U=\pi\left[ 
\begin{array}{ccc}
\frac{1}{\Omega \left( R+\rho \right) } & 0 & 0 \\ 
0 & \frac{1}{ \left(\Omega\rho +D\left( W_{v}
\right) \right)} & 0 \\ 
0 & 0 & \frac{1}{W_{v}}%
\end{array}%
\right] . \label{64}
\end{equation*}

The gain in sampling rates is given by the ratio%
\begin{equation}
\frac{\det \left( T\right) }{\det \left( U\right) }=\frac{4\left( R+\rho
\right) \left( \Omega \rho+D\left( W_{v}\right) \right) }{%
2R\Omega \rho +3RD+\rho D}.  \label{65}
\end{equation}
Let us assume that $\rho=c R$, for some constant $0<c<1$. Then, 
\begin{equation*}
\begin{split}
\frac{\det \left( T\right) }{\det \left( U\right) }&\leq \frac{4\left( R+\rho
\right) \left( \Omega \rho+D \right) }{2R\Omega \rho +(3R+\rho )D}
=\frac{4cR\Omega\left( 1+c
\right) +4(1+c)D}{2cR\Omega +(3+c )D}\\
&=1+\frac{cR\Omega\left( 2+c
\right) +(1+3c)D}{2cR\Omega +(3+c )D}.
\end{split}
\end{equation*}
For all practical parameters, the right hand term is smaller than 1, and thus the gain ratio is bounded by 2.

\begin{figure}[!]
  \centering
    \includegraphics[width=0.8\textwidth]{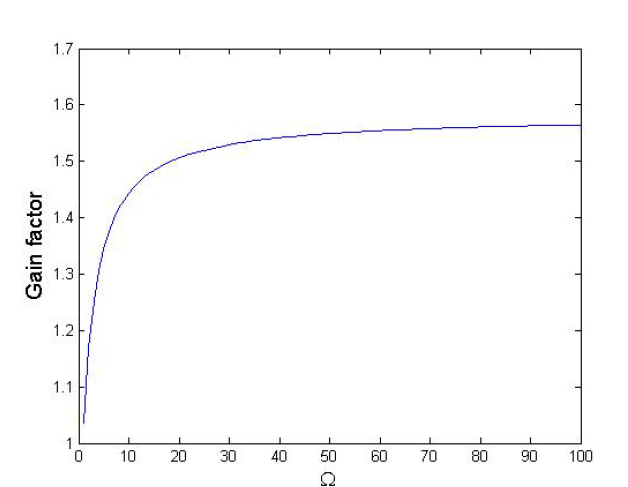}
  \caption{Sampling gain as a function of $\Omega$ according to (\ref{65}) with the parameters  R = 2[m],
Z = 0.4[m], h = 0.2[m], B = 2$\pi$Z/h [rad] and $\rho$= 0.5 [m].}
	\label{figGain}
\end{figure}

Consider now a CT system with fixed $R$ and $h$ and a scanned object of
dimensions $\rho $ and $2Z$. Then, the size of the essential support of the
scanned object represented by $\Omega $, determines the detail demands on
the image generated by the system. This is clearly translated into the
dimensions of the set defined in (\ref{59}). The larger $\Omega $ is the larger
this set becomes. This in turn means, finer sampling in the $\left( \alpha
,\beta ,v\right) $ space and a potential for a larger gain in a more
efficient sampling. This can readily be observed in the gain ratio as given
in equation (\ref{65}). To demonstrate this, we chose the set of parameters  $R = 2$[m],
$Z = 0.4$[m], $h = 0.2$[m], $B = 2\pi Z/h $[rad] and $\rho= 0.5$ [m], and present the resulting sampling rate gain as a function of $\Omega $ in Figure \ref{figGain}.

\section{Numerical Results}

\begin{figure}[!]
\begin{center}$
\begin{array}{ccrr}
\includegraphics[width=0.5\textwidth]{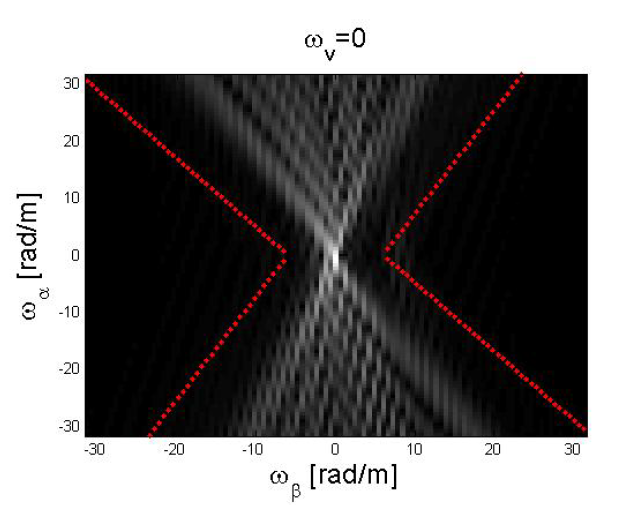} &
\includegraphics[width=0.5\textwidth]{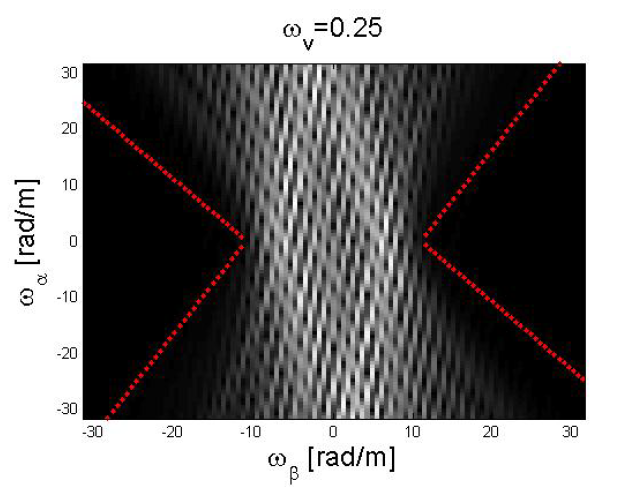} \\
\mbox{\bf (a)} & \mbox{\bf (b)} \\
\end{array}$
\end{center}
\caption{Numerical evaluation of $(\omega_\beta,\omega_\alpha$) plane according to equation (\ref{10}) for (a) $\omega_v=0$ (b) $\omega_v=0.5 [rad/sec]$. The red line is the set in equation (\ref{59}).}
\label{fig:support}
\end{figure}

Two numerical experiments were conducted. In the first experiment, we have numerically
calculated the integral in (\ref{10}) using the following values: R = 2.5[m],
Z = 1[m], h = 0.4[m], B = 2$\pi$Z/h [rad] and $\rho$= 0.5 [m]. In Figure \ref{fig:support}, we present
the results at $\omega_v = 0$ and $\omega_v =0.25 [rad/sec]$ and insert the bounds as in (\ref{59}). We clearly observe that
the calculated support is indeed contained in the set we derived and the shape
is also captured. However, our set seems to be quite conservatively evaluated.

The second experiment compared standard and efficient reconstructions of 3D Shepp-Logan phantom with the following values: R = 2[m],
Z = 0.4[m], h = 0.2[m], B = 2$\pi$Z/h [rad] and $\rho$= 0.5 [m]. 
The reconstruction from the proposed sampling was performed in two steps. First, the data was passed through a filter based on the support in equation
(\ref{59}). This stage can reduce noise effects.
 Then, a standard Katsevich reconstruction algorithm for curved detector was applied \cite{Katsevich,native_CB,rec_algorithm}. The reconstruction quality was assessed by three different criteria: visual assessment, mean square error (MSE) which computes the $\mathit{l}_2$ norm of the reconstruction error, and SSIM index which is based on the degradation of structural information, where SSIM index of 1 means a perfect match between two images \cite{ssim}. \\

\begin{figure}[!]
\begin{center}$
\begin{array}{ccc}
\includegraphics[width=0.5\textwidth]{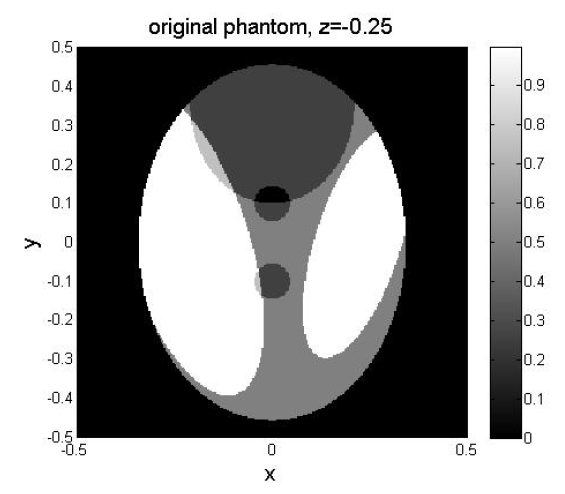} 
\end{array}$
\end{center}
\caption{ Slice of the original 3D phantom  }
\label{fig:slice}
\end{figure}

\begin{figure}[!]
\begin{center}$
\begin{array}{cc}
\includegraphics[width=0.5\textwidth]{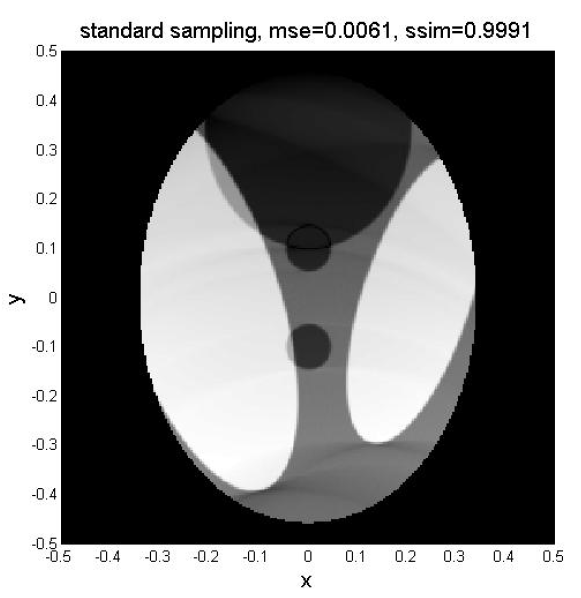} &
\includegraphics[width=0.5\textwidth]{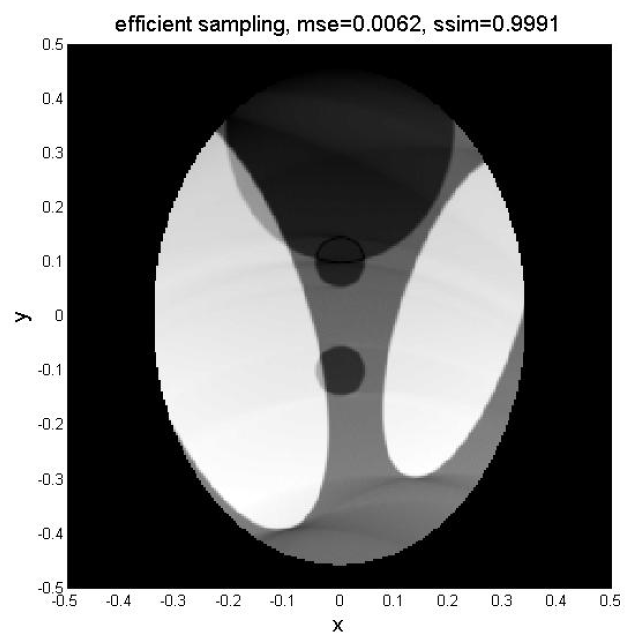} 
\\
\mbox{\bf (a)} & \mbox{\bf (b)} \\
\end{array}$
\end{center}
\caption{(a) reconstruction using standard sampling, with MSE=0.0061 and SSIM=0.9991, with respect to Figure \ref{fig:slice} (b) reconstruction using sparse  sampling,  with MSE=0.0062 and SSIM=0.9991, with respect to Figure \ref{fig:slice}.   }
\label{fig:reconstruction}
\end{figure}

In (\ref{fig:slice}), a region from a slice of the original 3D phantom is presented. 
We recall that our phantom is composed of ellipses for which the Fourier transforms are known, and thus its essential support can be estimated to be at most $3.3/a$, where $a$ is the smallest radius of the ellipse. In this experiment, $\Omega$ was chosen to be  $3.3/0.05=66[rad/m]$  \cite{analytical_magnetic_resonance_imaging_phantom}. 
Figure \ref{fig:reconstruction}(a) was reconstructed using standard (Nyquist) sampling, and Figure \ref{fig:reconstruction}(b) was reconstructed using the efficient sampling as in (\ref{61}).  The latter in this case has $35.77 \%$ less samples than the standard sampling. 

Only minor visual differences are observed in the reconstructed images. As expected, the standard sampled image shows a slightly better result compared with the sparse sampled image. The SSIM index of both reconstructions is identical, while the MSE of the standard sampling is $0.0061$ versus $0.0062$ of the sparse sampling.

\section{Conclusion and discussion}

The main result of this paper is the evaluation of the a bounding set in the
frequency domain for the HCB CT system. This result led to a more efficient
sampling pattern with gain ratio given in (\ref{65}). 
Furthermore, filtering the collected data with a low-pass filter according to (\ref{59}) can be used to attenuate the measurement noise. 

As a side result, we have calculated the Nyquist sampling rate of the scan in terms of its geometrical parameters and the essential support of the scanned object. This essential support can be easily calculated for several known shapes such as ellipsoid, cylinder and rectangular. Otherwise, it should be estimated based on past experience, trial and error or desired resolution. In Figure \ref{figGain}, we have shown that the larger $\Omega$ is, the larger the gain becomes.

We would like to conclude our paper with an idea which could suggest efficient sampling patterns for 
generalized HCB CT systems. As pointed out earlier in the paper, the rotation of
source/sensor array around the bed and the travel of the bed can be perceived
as a helical trajectory of the cone beam source on a cylindrical surface. By
viewing the position of the cone beam source as a point on a cylindrical surface
of radius $R$ with angle $\beta$ and height $\zeta$, the resulting scan becomes a 4D function
of $(\alpha,\beta,\zeta,v)$ which could be analyzed as such. One can then associate $\beta$ and
$\zeta$ resulting in a particular trajectory on the cylindrical surface in a way that
is better adapted to the scanned object. In particular, if $\zeta=\frac{h}{2\pi}\beta$ we have
the current scanning pattern. As an example, suppose now we have as a prior
information, that the scanned body has less detail in its middle than at its ends.
Then, a possible adaptation could be a variable pitch, smaller at the ends and
higher in the middle. This could lead to a more efficient sampling suited for this case. This idea is beyond the scope of the current paper but
definitely a topic for further research.


\appendix
\section*{{}Appendix A - Essential Supports of $\widehat{g}_{3}\left( \protect%
\omega _{\protect\beta }\right) $ and $\widehat{g}_{4}\left( \protect\omega %
_{\protect\beta },\protect\omega _{v}\right) $}
\setcounter{section}{1}

\bigskip Recall from (\ref{15}) that%
\begin{equation*}
\begin{split}
g_{3}\left( \beta \right)&=\frac{1}{1+\overline{r}^{2}-2\overline{r}\cos
\beta }, \\
g_{4}\left( \beta ,\omega _{v}\right) &=e^{-j2\omega _{v}\left( z-\frac{h}{%
2\pi }\beta \right) I_{B}\left( \beta \right) g_{5}\left( \beta -\phi
\right) } ,  \\
g_{5}\left( \beta \right) &=\frac{1-\overline{r}\cos \beta }{1+\overline{r}%
^{2}-2\overline{r}\cos \beta } . \label{A2}
\end{split}
\end{equation*}
Since both $g_{3}\left( \beta \right) $ and $g_{5}\left( \beta \right) $ are
periodic in $\beta $, we have 
\begin{align*}
\widehat{g}_{3}\left( \omega _{\beta }\right) &=\sum_{k\in \mathbb{Z}}%
\widehat{g}_{3,k}\delta \left( \omega _{\beta }-k\right). \\
\widehat{g}_{5}\left( \omega _{\beta }\right) &=\sum_{n\in \mathbb{Z}}%
\widehat{g}_{5,n}\delta \left( \omega _{\beta }-n\right),
\end{align*}
where, for $k\geq 0$%
\begin{equation*}
\widehat{g}_{3,k} =\frac{1}{2\pi }\int_{-\pi }^{\pi }\frac{e^{-jk\beta }}{%
1+\overline{r}^{2}-2\overline{r}\cos \beta }d\beta 
=\frac{1}{2j\pi \overline{r}}\oint \frac{z^{k}dz}{z^{2}-\frac{1+\overline{%
r}^{2}}{\overline{r}}z+1} \\
\end{equation*}%
Since $0\leq \overline{r}<1$ we can use the residue theorem to get%
\begin{equation*}
\widehat{g}_{3,k}=\frac{\overline{r}^{k}}{\overline{r}^{2}-1},
\end{equation*}%
and as $\widehat{g}_{3,k}\left( k\right) =$ $\overline{\widehat{g}_{3,k}\left(
-k\right) }$ we conclude that%
\begin{equation*}
\left\vert \widehat{g}_{3,k}\right\vert =\frac{\overline{r}^{\left\vert
k\right\vert }}{1-\overline{r}^{2}}\text{ \ for all }k . \label{A3}
\end{equation*}%
Then, using the definition of the essential support in (\ref{6}), we
conclude that%
\begin{equation*}
\mathcal{L}\left\{ \widehat{g}_{3}\left( \omega _{\beta }\right) \right\}
=\left\{ \omega _{\beta }:\left\vert \omega _{\beta }\right\vert \leq
K\right\},  \label{A4}
\end{equation*}%
where%
\begin{equation}
K=\left\lfloor \frac{\log \left( 0.01\left( 1+\overline{r}^{2}\right)
\right) }{2\log \overline{r}}\right\rfloor.  \label{A5}
\end{equation}%
Similarly, we get%
\begin{equation*}
\begin{split}
\widehat{g}_{5,n} &=\frac{1}{2\pi }\int_{-\pi }^{\pi }\frac{\left( 1-%
\overline{r}\cos \beta \right) e^{-jn\beta }}{1+\overline{r}^{2}-2\overline{r%
}\cos \beta }d\beta \\
&=\frac{1}{4j\pi \overline{r}}\oint \frac{\left( 2z-\overline{r}\left(
z^{2}+1\right) \right) z^{n}dz}{z\left( z-\overline{r}\right) \left( z-\frac{%
1}{\overline{r}}\right) },
\end{split}
\end{equation*}
and again, from the residue theorem we conclude%
\begin{equation*}
\widehat{g}_{5,n}=\left\{ 
\begin{array}{l}
-1\text{\ for }n=0 ,\\ 
-\frac{1}{2}\overline{r}^{\left\vert n\right\vert }\text{ \ for }n\neq 0.%
\end{array}%
\right.  \label{A6}
\end{equation*}%
So, again%
\begin{equation}
\mathcal{L}\left\{ \widehat{g}_{5}\left( \omega _{\beta }\right) \right\}
=\left\{ \omega _{\beta }:\left\vert \omega _{\beta }\right\vert \leq
K\right\} . \label{A7}
\end{equation}%
$K$ as given in (\ref{A5}).

Before tackling the essential support of $g_{4}\left( \beta ,\omega
_{v}\right) $, we need to look at the Fourier transform of $\beta I_{B}\left(
\beta \right) g_{5}\left( \beta \right) $. Since%
\begin{equation*}
\int_{-\infty }^{\infty }\beta I_{B}\left( \beta \right) e^{-j\beta \omega
_{\beta }}d\beta =\frac{2j}{\omega _{\beta }^{2}}\left( \omega _{\beta
}B\cos \left( B\omega _{\beta }\right) -\sin \left( B\omega _{\beta }\right)
\right),  \label{A7a}
\end{equation*}%
using Fourier transform properties we get%
\begin{equation*}
\int_{-\infty }^{\infty }\beta I_{B}\left( \beta \right) g_{5}\left( \beta
\right) e^{-j\beta \omega _{\beta }}d\beta =\left( \int_{-\infty }^{\infty
}\beta I_{B}\left( \beta \right) e^{-j\beta \omega _{\beta }}d\beta \right)
\ast \widehat{g}_{5}\left( \omega _{\beta }\right).  \label{A8}
\end{equation*}%
Hence,
\begin{equation}
\mathcal{L}\left\{ \int_{-\infty }^{\infty }\beta I_{B}\left( \beta \right)
g_{5}\left( \beta \right) e^{-j\beta \omega _{\beta }}d\beta \right\} =%
\mathcal{L}\left\{ \widehat{g}_{5}\left( \omega _{\beta }\right) \right\} +%
\mathcal{L}\left\{ \frac{2j}{\omega _{\beta }^{2}}\left( \omega _{\beta
}B\cos \left( B\omega _{\beta }\right) -\sin \left( B\omega _{\beta }\right)
\right) \right\} . \label{A9}
\end{equation}%
To find the second term in (\ref{A9}), we use Definition \ref{definition2}. As%
\begin{equation*}
\int_{-\infty }^{\infty }\left\vert \frac{2j}{\omega _{\beta }^{2}}\left(
\omega _{\beta }B\cos \left( B\omega _{\beta }\right) -\sin \left( B\omega
_{\beta }\right) \right) \right\vert ^{2}d\omega _{\beta }=\frac{4B^{3}\pi }{%
3},
\end{equation*}%
and%
\begin{eqnarray*}
&&\int_{-W}^{W}\left\vert \frac{2j}{\omega _{\beta }^{2}}\left( \omega
_{\beta }B\cos \left( B\omega _{\beta }\right) -\sin \left( B\omega _{\beta
}\right) \right) \right\vert ^{2}d\omega _{\beta } \\
&=&4B^{3}\left( \frac{2}{3}\int_{0}^{2WB}\frac{\sin x}{x}dx+\frac{2WB\sin
\left( 2WB\right) +\left( 1+W^{2}B^{2}\right) \cos \left( 2WB\right)
-1-3W^{2}B^{2}}{3\left( WB\right) ^{3}}\right)
\end{eqnarray*}%
by trial and error we found that 
\begin{equation}
\mathcal{L}\left\{ \frac{2j}{\omega _{\beta }^{2}}\left( \omega _{\beta
}B\cos \left( B\omega _{\beta }\right) -\sin \left( B\omega _{\beta }\right)
\right) \right\} =\left\{ \omega _{\beta }:\left\vert \omega _{\beta
}\right\vert \leq \frac{48}{B}\right\} . \label{A10}
\end{equation}%
Hence, by (\ref{A7}), (\ref{A9}) and (\ref{A10}) we get%
\begin{equation*}
\mathcal{L}\left\{ \int_{-\infty }^{\infty }\beta I_{B}\left( \beta \right)
g_{5}\left( \beta \right) e^{-j\beta \omega _{\beta }}d\beta \right\}
=\left\{ \omega _{\beta }:\left\vert \omega _{\beta }\right\vert \leq K+%
\frac{48}{B}\right\}.  \label{A11}
\end{equation*}

To find constraints on the essential support of $\widehat{g}_{4}\left(
\omega _{\beta },\omega _{v}\right) $ we rewrite $g_{4}\left( \beta ,\omega
_{v}\right) $ as%
\begin{equation*}
g_{4}\left( \beta ,\omega _{v}\right) =e^{-j2\omega _{v}\left( z-\frac{h}{%
2\pi }\beta \right) I_{B}\left( \beta \right) g_{5}\left( \beta -\phi
\right) }  
=e^{-jC\left( z,\omega _{v}\right) g_6\left( \beta \right) } , \label{A12}
\end{equation*}%
where%
\begin{equation*}
C\left( z,\omega _{v}\right) =\frac{2\left\vert \omega _{v}\right\vert }{1-%
\overline{r}}\left( \left\vert z\right\vert +\frac{h}{2\pi }B\right),
\label{A13}
\end{equation*}%
and%
\begin{equation*}
g_6\left( \beta \right) =\frac{2\omega _{v}\left( z-\frac{h}{2\pi }\beta
\right) I_{B}\left( \beta \right) g_{5}\left( \beta -\phi \right) }{C\left(
z,\omega _{v}\right) } . \label{A14}
\end{equation*}%
Since $\left\vert g_{5}\left( \beta -\phi \right) \right\vert <\frac{1}{1-%
\overline{r}}$, we readily observe that $\left\vert g_6\left( \beta \right)
\right\vert \leq 1$. Then, we use the approximation%
\begin{equation*}
e^{-jC\left( z,\omega _{v}\right) g_6\left( \beta \right) }\approx
e^{-jC\left( z,\omega _{v}\right) \sin \left( \left( K+\frac{48}{B}\right)
\beta \right) } , \label{A15}
\end{equation*}%
to apply Carson's rule to determine the constraint for the essential support
of $\widehat{g}_{4}\left( \omega _{\beta },\omega _{v}\right) $.
Specifically, Carson's rule says that if a signal can be written as \cite{Principles_of_Communication_Systems}%
\begin{equation}
\cos \left( \left( \frac{\Delta \omega }{\omega _{m}}\right) \sin \left(
\omega _{m}\beta \right) \right)  \label{A16}
\end{equation}%
then its essential support is contained in $\left\{ \omega _{\beta
}:\left\vert \omega _{\beta }\right\vert \leq \Delta \omega +\omega
_{m}\right\} $. Comparing (\ref{A15}) and (\ref{A16}). we conclude that%
\begin{equation*}
\mathcal{L}\left\{ \widehat{g}_{4}\left( \omega _{\beta },\omega _{v}\right)
\right\} \subset \left\{ \omega _{\beta }:\left\vert \omega _{\beta
}\right\vert \leq \left( K+\frac{48}{B}\right) \left( 1+C\left( z,\omega
_{v}\right) \right) \right\} . \label{17}
\end{equation*}

\end{document}